# Community Aware Random Walk for Network Embedding

Mohammad Mehdi Keikha[1, 2], Maseud Rahgozar[1], Masoud Asadpour[1]

Email: {mehdi.keikha, rahgozar, asadpour} @ut.ac.ir

Corresponding Author: Maseud Rahgozar

**Abstract:**

Social network analysis provides meaningful information about behavior of network members that can be used for diverse applications such as classification, link prediction. However, network analysis is computationally expensive because of feature learning for different applications. In recent years, many researches have focused on feature learning methods in social networks. Network embedding represents the network in a lower dimensional representation space with the same properties which presents a compressed representation of the network. In this paper, we introduce a novel algorithm named "CARE" for network embedding that can be used for different types of networks including weighted, directed and complex. Current methods try to preserve local neighborhood information of nodes, whereas the proposed method utilizes local neighborhood and community information of network nodes to cover both local and global structure of social networks. CARE builds customized paths, which are consisted of local and global structure of network nodes, as a basis for network embedding and uses the Skip-gram model to learn representation vector of nodes. Subsequently, stochastic gradient descent is applied to optimize our objective function and learn the final representation of nodes. Our method can be scalable when new nodes are appended to network without information loss. Parallelize generation of customized random walks is also used for speeding up CARE.

We evaluate the performance of CARE on multi label classification and link prediction tasks. Experimental results on various networks indicate that the proposed method outperforms others in both Micro and Macro-f1 measures for different size of training data.

## 1. Introduction:

There has been remarkable growth in online social networks and the number of their users. Valuable information can be extracted from social networks by analyzing both their structure and content. Machine learning techniques are used as a way to extract valuable features from social networks for different analysis tasks such as classification [1, 2, 3], recommendation [4, 5] and link prediction [6, 7, 8, 9]. These learning methods can be both supervised and unsupervised. Supervised learning algorithms are able to extract features better for a specific task on social networks but their scalability would be challenging for large networks. On the other hand, unsupervised methods can handle scalability of feature learning methods; however, the extracted features show low accuracy in different network analysis tasks. They are too general to give valuable information for a specific task. [10, 11, 12, 13, 14, 15, 16].

---

[1] School of Electrical and Computer Engineering, College of Engineering, University of Tehran, Tehran, Iran
[2] University of Sistan and Baluchestan, Zahedan, Iran

Network embedding, as an unsupervised representation learning task, tries to extract informative lower dimensional representation of network nodes. It learns social relationships of network nodes in a low dimensional space to preserve both microscopic and macroscopic network structure including various proximity orders, community membership and their inherent properties. These representation vectors can be used in different social network analysis tasks such as classification [17], recommendation [18] and link prediction [6]. Some of classic network embedding methods use eigenvectors of affinity graph as feature vectors [10, 15, 19, 20]. Graph factorization is another technique which is used for network embedding [21]. The aforementioned approaches suffer from scalability for large social networks.

In recent years, deep learning as an unsupervised method is widely used in natural language processing which a detailed description of these researches can be found in [11]. There are also many researches that have used deep learning for social network embedding [22, 23, 24, 25]. Network embedding methods try to represent graph nodes with some informative feature vectors. DeepWalk [22], LINE [23] and Node2vec [24] are the most important methods that are proposed in the recent years. Though, these methods show good performance in comparison to other graph representation methods such as Spectural clustering, but they attempt to extract only local structural information from each node, and then employ them to learn final representation of the node. However, communities are important structural information ignored by these methods [26].

Community structure imposes constraints in a higher structural level on the nodes' representation. The representation of nodes within a community should be more similar than those belonging to different communities. Furthermore, for two nodes within a community, even if they only have weak relationship in local structure due to the data sparsity issue, their similarities will also be strengthened by the community structure constraint. Thus, incorporating community structure in network embedding can provide effective and rich information to solve data sparsity issues in global structures and moreover, make the learned nodes' representation more discriminative [25].

In this paper, we propose a new network embedding method called "CARE," which utilizes community information of network nodes to capture more structural information of networks. Some previous researches tried to embed community information on nodes' representation. For instance, Grover et al. in [24] only consider the community members that their distance to the source nodes is less than 2. However, in real-world networks which communities have thousands of members, Node2vec would not be able to consider information about the nodes that their distance is more than two from the source of random walk because Node2vec creates second order random walks. CARE can also produce the representation vector of nodes for arbitrary type of networks such as weighted, complex and directed.

CARE, firstly, extracts communities of the input network. We prepare this information with the Louvain method [27] which has effective performance on different social networks. To learn final representations, we generate some community aware random walks that consider both first and higher order proximities as well as community membership information for each node. The customized paths contain the nodes that are in the same neighborhood structure as well as nodes that belong to the same community. CARE makes several customized paths for each network node to embed different structural information into final representation. Finally, the customized random walks are used as contextual information to learn final representation of nodes by the Skip-gram learning model.

CARE is evaluated with two social network analysis tasks: multi label classification and link prediction. The experimental results show that CARE outperforms Node2vec with a gain of 50% on multi label classification with BlogCatalog dataset and 3% on the link prediction task for PPI dataset.

To summarize, we make the following contributions:

- We present a novel network embedding algorithm named CARE that learns the representation of nodes for different types of networks such as: weighted, directed and complex networks.
- Our method can preserve community information of the network in the learned representation vectors while the previous researches are not able to define an optimization function considering this information explicitly.
- CARE preserves all properties of the network structure through the generation of customized paths for each node, independently. Therefore, it spends less time to learn final representations of nodes because of parallel path generation.
- We empirically evaluate the algorithm on multi label classification and link prediction problems with different real world social networks. The experimental results indicate the efficiency of CARE in contrast to other network embedding methods.

The rest of paper is organized as follows: In section 2, we summarize related works to network embedding. We explain details of CARE in section 3. Section 4 outlines the experimental results on two network analysis tasks. Finally, Section 5 presents conclusion and future works.

## 2. Related Works:

In this section, we review recent researches related to unsupervised representation learning of network nodes. Some feature learning approaches use adjacency matrix of the network and try to preserve the first order proximity of nodes. These researches act as dimension reduction methods and find the best eigenvectors of network matrices [10, 15, 16, 19, 20, 21, 28] to use as the feature vector of networks. Eigenvector decomposition is usually computationally expensive. Furthermore, they only consider immediate neighborhood of nodes and do not use higher order proximities and community information. So, they are unable to preserve the global structure of networks. As a result, the learned representations would not provide an appropriate performance on diverse network analysis tasks.

In recent years, deep learning is used as an alternative to learn feature vector of network nodes. These methods have utilized deep learning to learn representation vectors. They generate random walks with different graph exploration strategies and have embedded them as contextual information into the Skip-gram model. DeepWalk was the first method that used the Skip-gram model [22]. It treats DFS like search strategy to generate random walk. Despite the good performance on multi label classification, this method failed to preserve global network structure because it does not consider community information of network nodes. LINE uses first and second order proximities to learn nodes' representation, but it also preserves local information of the networks [23]. The authors in [23] define two independent functions for first and second order proximities but they ignore community information. LINE and DeepWalk also fail to learn representation vector for network edges.

Node2Vec makes random walks based on DFS and BFS like strategies [24]. While Node2vec uses two controlled parameters to consider both homophily [29] and structural equivalences [30] of networks, it does not guarantee to reach different nodes of a community. The main reason for this problem is that these algorithms only consider second order proximities and cannot reach the nodes that their distance is more than 2 from the start node of random walk. Because in real networks, there are many nodes in a community and obviously their distance is greater than two, thus Node2vec would not consider all the community members during creation of random walks for a node. SDNE proposes a semi-supervised deep model, which has multiple layers of non-linear functions, thereby being able to capture the highly non-linear network structure [31]. It exploits the first and second-order proximity jointly to preserve the network structure, but it doesn't use community information.

The proposed method in [25] uses modularized non negative matrix factorization to preserve both microscopic and macroscopic information of networks. The authors in [25] define two independent model to embed local and community information independently and then optimize the joint function to learn the representation of nodes. They learn local and community structure separately. Consequently, they combine the final representations. Their final representation is not general enough to be used in different network analysis tasks because It also has some local structure information loss because it combines first and second order proximities in a unified matrix. Unification of matrices leads to missing information about different proximities during representation learning. Their method also suffers from scalability when the networks are large because they should learn many parameters to preserve local and global structures, thus it is not applicable on real social networks.

Unlike previous researches, we employed a mixture of BFS and DFS like strategies alongside community information of network nodes without any restriction of search length over search space. We preserve both local and global information because we use first and higher order proximities as well as community information of nodes to learn nodes' representations.

## 3. CARE: Community Aware Random Walk for Network Embedding

Community information is one of the key features of social networks, which preserves the global structure of the network [26]. However, it is ignored by the most previous researches in network embedding when they want to gather information about network nodes. We present a new algorithm to embed graph structure alongside community information into the learned representation vectors of network nodes. Therefore, we redefine network embedding as a maximum likelihood problem which is gained by global network structures. Suppose G = (V, E) is an (un) directed graph which V and E are set of graph nodes and edges. We are going to find a mapping function f: V → $R^d$ which d is the representation size of each graph node. To obtain the best mapping function f, the Skip-gram model is used [32, 33].

In CARE, first neighborhood structure for each node is extracted from the given network using community aware random walk strategy. Subsequently, by using the Skip-gram model, the representation vector of the node is learned from these generated random walks. Most of the previous approaches for modeling neighborhood structure of a node have only used the first and the second order proximities. In contrast, we use the nodes that may not have an immediate connection or second order proximity with the source node. However, they have a homophily relationship which is not presented by the first and the second order proximities.

Once different neighborhood structures are extracted for all nodes, we use the Skip-gram model similar to [24] to maximize N(u); where N(u) contains the neighborhood structure of a node u. The Skip-gram model learns the best representation vector for the node u based on the structural information contained in N(u). In the following, we explain how we create neighborhood structure and how they are used to learn social representations of a node in the given network. Algorithm 1 illustrates the steps of CARE algorithm.

---

Algorithm 1: CARE (G, $w$, d, µ, l)

---

**Input**:

    Graph G (V, E)

    Window length $w$

    Representation size d

    Number of random walks per node µ

    Random walk max length l

**Output**:

    Matrix of node representations $f \in R^{|v| \times d}$

1: Com = CommunityDetection (G)

2: sample $f$ from $U^{|v| \times d}$

3: while (i < µ)

4:     S = shuffle (V)

5:     for each $v_i \in$ S do

6:         $\mathcal{W}_{v_i}$ = CommunityawareRW ( G, $v_i$ , Com, l )

7:         SkipGram ($f$, $\mathcal{W}_{v_i}$ , $w$)

8:     end for

9: end while

---

In algorithm 1, Line 1 detects the communities of the given graph G. The Louvain method is used for detecting communities, which is explained in section 3.1. Before we learn the optimal representation vectors for graph nodes, we generate the matrix U randomly to initialize the representation vector of nodes in line 2. Now we are able to learn final representation vectors in lines 3-9 of Algorithm 1. For each node in V, it is generated µ different customized random walks to better capture the global and local structure of the node in line 4. Before iterating over network nodes, they are shuffled to avoid the effect of nodes visiting order in the final representations. The core of the presented method for network embedding is line 6 where we generate customized random walks for the chosen node which would be clarified in section 3.2. Finally, the generated paths are used to update node representation in line 7. In the following sections, different functions of Algorithm 1 are explained in more details.

### 3.1. Community Detection:

We have used the Louvain method to maximize modularity in the network to detect communities [27]. Modularity is a metric to compare the density of edges that are inside a community to the edges between communities. It is an optimization algorithm that firstly considers each node in a separate community. Subsequently, a node is chosen and the modularity of joining the node to the neighbors' communities are calculated. It finally assigns the node to the community, in which the modularity is maximized. Modularity in the network is calculated using the following formula:

$$Q = \frac{1}{2m} \sum_{ij}[A_{ij} - \frac{k_i k_j}{2m}] \; \delta \, (c_i . c_j) \qquad (1)$$

In Eq. 1, m stands for sum of all edge weights in the graph. $A_{ij}$ denotes the edge weight between nodes i and j. Summation of edge weights of i and j are represented by $k_i . k_j$. The communities of i and j are shown by $c_i$ and $c_j$. Finally $\delta$ is a delta function that returned 1 when communities are equal.

As the modularity maximization problem is intractable, we have first used a heuristic version of the Louvain method to find communities, then each small community is considered as a node in a new network, and we try to maximize the modularity with the new network [27].

### 3.2. Generation of neighborhood structure:

To extract neighborhood structure of a node, we build μ customized random walks. A customized random walk that starts from node v is shown by $\mathcal{W}_v$. Since a random walk is a path in the given network; we can denote a customized random walk for node v with some random variables $\mathcal{W}_v^1$, $\mathcal{W}_v^2$, ..., $\mathcal{W}_v^k$ such that $\mathcal{W}_v^{k+1}$ is a node selected at random from immediate neighbors or the nodes that are in a same community with k-th node of the path.

To create a customized random walk started from node v, we first extract all of its immediate neighbors. Then a random variable r between 0 and 1 is generated. If r is less than α, we pick a node at random from the immediate neighbors; otherwise, we choose a node from the nodes that are in the same community with k-th node of $\mathcal{W}_v$ as it is shown in Eq.2.

$$\mathcal{W}_v^{k+1} = \begin{cases} immediate\ neighbors & 0 < r \leq \alpha \\ nodes\ in\ community\ of\ k-th\ node & \alpha < r < 1 \end{cases} \qquad (2)$$

If k-th node of $\mathcal{W}_v$ be a member of several communities, we first extract all the members of these communities. Then we choose one of them, randomly.

This process is continued until it reaches a predefined length l for the path. Furthermore, if a node in the path has no new neighbor, we stop expanding the path. Algorithm 2 explains the details of forming customized random walks in CARE.

---

Algorithm 2: CommunityawareRW ( G, $v_i$, $Com_{v_i}$, l, α)

---

**Input**:

    Graph G (V, E)

    Source node of RW $v_i$

    Nodes belong to the same community with $v_i$ $Com_{v_i}$

    Random walk max length l

|  |
|---|
| Random variable to select from neighbors or same community members α |

**Output**:

    A path with max length l

---

1: initialize RW with $v_i$

2: while length(path) < l

3:      if current node has neighbors

4:          if (random (0, 1) < α)

5:              select $v_j$ at random from $v_i's$ neighbors

6:          else

7:              select $v_j$ at random from members of $v_i's$ communities

8:      else

9:          backtrack in the path and select the last node which has neighbors that are not in the path

10: end while

---

The proposed random walk generator extracts local and global information from the given network. In addition, we can parallelize the algorithm to speed up the process of network embedding. This is because the customized random walks are made independently from each other. Additionally, if some new nodes are added to (removed from) the network, their biased random walks are generated without the need to obtain new customized random walks for previous nodes.

### 3.3. Skip-Gram:

According to Algorithm 1, after generation of random walks, we use the Skip-gram model to learn representation of graph nodes [32, 33]. The Skip-gram is a language model that maximizes conditional probability of words' co-occurrence in a predefined window $w$ as it is shown in Eq.3:

$$\Pr(w \mid f(u)) = \max_f \prod_{\substack{j=i-w \\ j \neq i}}^{i+w} \Pr(v_j \mid f(u)) \quad w = \{v_{i-w}, \dots, v_{i+w}\} \backslash u \quad (3)$$

For each node in the given network, we iterate over all its customized paths. We define a window $w$ to slide over a path. Similar to the previous approaches, the independence assumption of conditional probabilities is considered in Eq. 3. Additionally, Softmax function is used to approximate the probability distribution of Eq. 3 as the following in Eq. 4:

$$\Pr(v_j \mid f(u)) = 1 / (1 + e^{-f(u).f(v_j)}) \quad (4)$$

Furthermore, Stochastic gradient descent (SGD) is used to optimize the parameters similar to the proposed method in [34]. In the beginning of the training, the learning rate is 2.5%, and it decreases linearly with the number of vertices that are seen so far as stated in [22].

For complex networks, we consider different edges of two nodes independently. If there are more edges between nodes in the given network, it is more probable to choose these nodes along the path. If the given network is weighted, we consider weights of edges as a probability to pick the edges when producing customized random walks.

In large scale networks, CARE prepared different communities with the Louvain method in parallel settings. The customized paths are obtained by multiple threads simultaneously, since the path generation for each node is done independently.

Our algorithm is also scalable when some new nodes are appended to (removed from) the network. Customized random walks are only generated for new nodes, and their representation vectors are calculated as stated above. We are able to generate the customized random walks in parallel to increase the speed of CARE. The time complexity of CARE is the same as the Skip-gram model. When the community detection and path generation phases are finished, the representation learning of nodes is started. The biggest time complexity among these phases corresponds to node representation learning of the Skip-gram model.

## 4. Experiments:

In this section, we evaluate CARE with two supervised learning tasks: multi label classification and link prediction. We also analyze the effect of different parameters. We compare our results on the aforementioned tasks with the best representation learning methods, which are explained in section 4.1.

### 4.1. Baseline Algorithms:

To evaluate the performance of the proposed algorithm, we compare it with the following representation learning algorithms that have the best results on multi label classification and link prediction tasks:

- **Spectral clustering** [28]: This algorithm attempts to find graph cuts that lead to better classification of the graph. Therefore, it first calculates the normalized Laplacian matrix of the graph G. Then, it considers the d-smallest eigenvectors of the matrix as the best feature vector to represent graph nodes.
- **DeepWalk** [22]: DeepWalk is the first algorithm that uses deep learning for social network embedding. It generates random walks to learn representation vector of the nodes. DeepWalk could be considered as a variant of CARE with $\alpha = 0$ in algorithm 2.
- **LINE** [23]: LINE uses local information, including first and second order proximities of nodes instead of generating random walks. It, firstly, defines two separate functions to preserve immediate relations and second order proximities in a social network. In the second stage, two functions are combined linearly to calculate the final representation of each node.
- **Node2vec** [24]: Node2vec is a semi supervised algorithm that generates a second order random walk to capture network neighborhood information of nodes. It uses two parameters to simulate BFS and DFS search strategies.
- **M-NMF** [25]: This algorithm learns final representation of nodes using two independent functions that generate two matrices. The first matrix keeps information about the local structure of network, including the first and second order proximities, while the second matrix contains the representation of network communities. Finally, the Non negative matrix factorization is used to preserve all the network structural information.

**Parameter settings**: To compare our results with the above algorithms, we have used the same parameter settings that are reported in [24] for all the algorithms. We set $w$= 10, $d$= 128, $\mu$= 10, $l$=80. The optimal value for $\alpha$ is 0.2. We have employed the same datasets and experimental procedure as [24]. The best values for p and q in Node2vec algorithm are chosen from {0:25, 0:5, 1, 2, 4} as stated in [24].

### 4.2. Multi label classification:

In the multi label classification task, we predict one or more label for each network node. To compare our algorithm with baseline algorithms, we evaluate the methods with the following datasets:

**BlogCatalog** [35]: This is a social network of bloggers in which, the node's labels are topic categories generated by each blogger. It has 10312 nodes and 333983 edges and 39 different topic labels.

**Protein-Protein Interactions (PPI)** [36]: This is a subgraph of Homo Sapiens PPI network which is preprocessed in [24], including 3890 nodes, 76584 edges and 50 labels which are extracted from gene sets.

**Wikipedia** [37]: It is a co-occurrence words' network of Wikipedia's articles that has 4777 nodes, 184812 edges and 40 different labels. The labels of nodes are part of speech (POS) tags of network nodes.

Table 1 summarizes statistics of datasets that are used in the multi label classification task.

|  | |V| | |E| | Labels |
|---|---|---|---|
| BlogCatalog | 10312 | 333983 | 39 |
| Protein-Protein Interactions (PPI) | 3890 | 76584 | 50 |
| Wikipedia | 4777 | 184812 | 40 |

*Table 1 datasets that used in multi label classification*

To learn a classifier, we use a fraction of the learned representations along with their labels as training set. The rest of the nodes' representations are used to evaluate the performance of all algorithms in multi label classification. The regression classifier is used to predict the nodes' labels of the test set.

### 4.2.1. Experimental Results:

In the experiments, the training size of input datasets is increased from 10% to 90%. Micro and Macro-f1 measures are applied to evaluate performance of different algorithms [31]. Micro-f1 considers equal weights to each data instance, while Macro-f1 is a metric which gives equal weights to each class. They are defined as follows:

$$Precision = \frac{\sum_{A \in C} TP(A)}{\sum_{A \in C}(TP(A) + FP(A))} \qquad (5)$$

$$Recall = \frac{\sum_{A \in C} TP(A)}{\sum_{A \in C}(TP(A) + FN(A))} \qquad (6)$$

$$Micro - f1 = \frac{2 * Precision * Recall}{Precision + Recall} \qquad (7)$$

$$Macro - f1 = \frac{\sum_{A \in C} Micro - f1\ (A)}{|C|} \tag{8}$$

In aforementioned formulas, TP(A), FP(A) and FN(A) are the number of true positives, false positives and false negatives in the instances which are predicted as A, respectively. Suppose C is the overall label set. Micro-f1(A) is the Micro-f1 measure for the label A.

Figure 1 shows the performance of CARE in comparison to the other methods for multi label classification task over different networks. In the following, we discuss the experimental results for each dataset.

- **BlogCatalog:**

For BlogCatalog dataset, our method shows significant improvements over both Micro and Macro-f1. When there are only 10% training data, CARE achieves a gain of 50% over Node2vec. In this condition, both LINE and Spectral clustering show poor performance because they only use local neighborhood information. M-NMF does not also provide good performance than deep learning based methods because it combines first and second order proximity in a unified matrix. Thus, network sparsity cannot be handled by these algorithms. This property would be useful, especially in sparse networks. The most important difference between CARE and the other algorithms is the usage of community information during path generation. As a result, we will be able to maintain both local and global structural information of the network when learning the representation vector of each node.

- **PPI**:

As the results shows in figure 1, both evaluation metrics have less value than BlogCatalog network because PPI network has lower density in comparison to BlogCatalog. CARE has significant improvements over Node2vec, DeepWalk and Community preserving methods. When training data is 50%, our method outperforms about 50 % over Node2vec because we utilize community membership information during nodes' representation learning. M-NMF also gives weak results in the dataset because it suffers from local structure information loss as discussed in section 2.

- **Wikipedia**:

As another evaluation, we test CARE on the co-occurrence word network of Wikipedia's articles. The results of experiments show CARE outperformed baseline algorithms, considering both Micro and Macro-f1. When training data size reaches 80%, our method achieved the highest improvements of 7% over Node2vec. Wikipedia dataset is denser than the other datasets. So, M-NMF cannot predict nodes' labels because it may not preserve local information directly during the learning of nodes' representation.

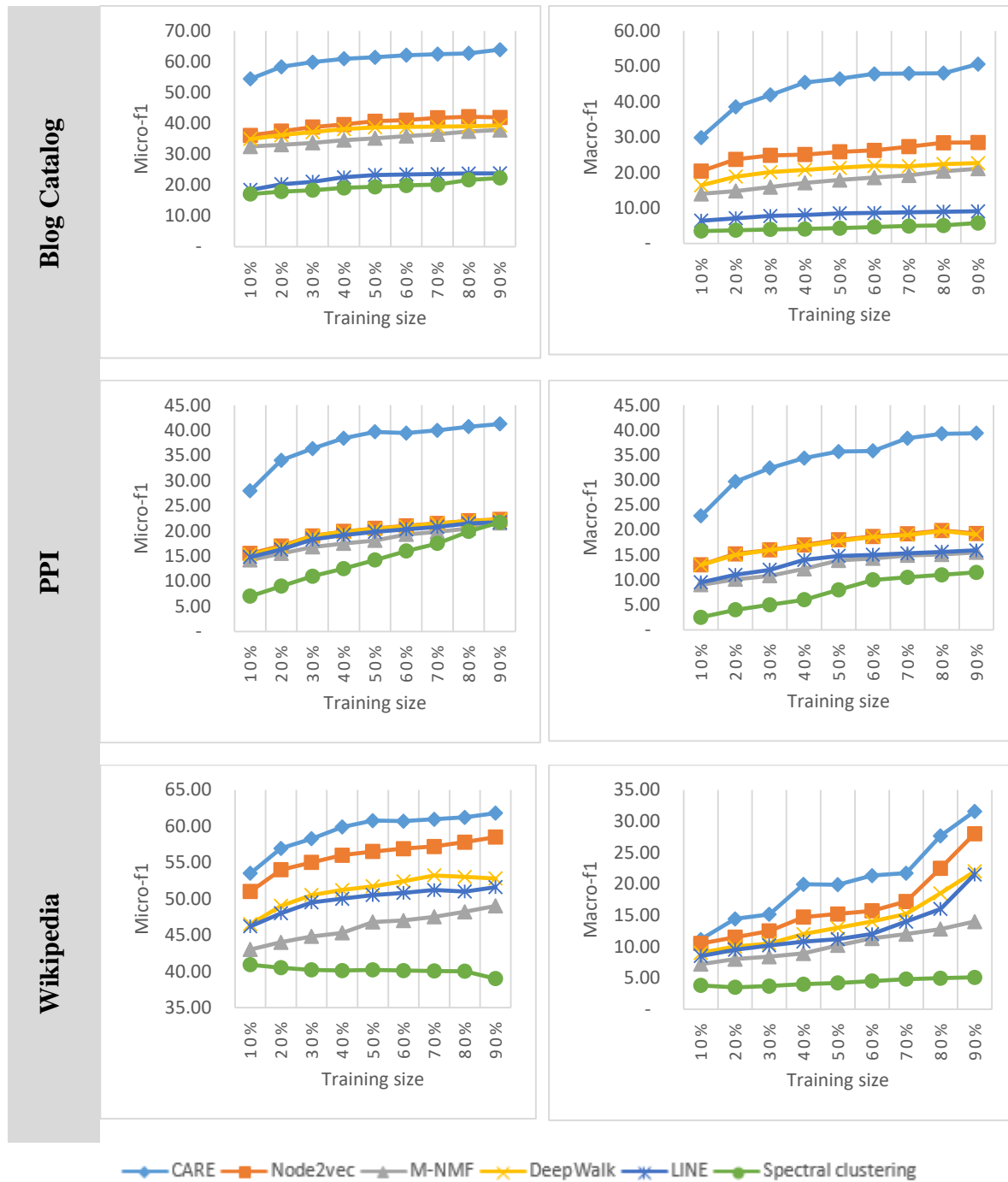

*Figure 1 Micro and Macro-f1 scores on Multi label classification task*

As it has shown in figure 1, when the network is sparse, there is less local information for nodes and as a result, most networks embedding methods cannot perform the multi label classification task well. The average Micro and Macro-f1 measures for different datasets are also reported in Table 2.

|  | BlogCatalog | | PPI | | WikiPedia | |
|---|---|---|---|---|---|---|
| Algorithm | Micro-f1 | Macro-f1 | Micro-f1 | Macro-f1 | Micro-f1 | Macro-f1 |
| CARE | 60.69 | 44.09 | 37.55 | 34.21 | 59.33 | 20.30 |
| Node2vec | 39.98 | 25.61 | 19.86 | 17.37 | 55.88 | 16.42 |
| DeepWalk | 37.94 | 20.70 | 19.77 | 17.23 | 51.14 | 13.80 |
| M-NMF | 35.2 | 17.66 | 18.14 | 12.86 | 46.17 | 10.31 |
| LINE | 22.27 | 8.15 | 19.19 | 13.68 | 49.87 | 12.63 |
| Spectural Clustering | 19.57 | 4.45 | 14.31 | 7.61 | 40.12 | 4.28 |

*Table 2 Average Micro and Macro-f1 for different datasets*

### 4.2.2. Parameter sensitivity:

In this experiment, the best parameter values for CARE are found on BlogCatalog in the multi label classification task. In each experiment, we consider default values for all the parameters and change only one parameter. We also pick 50% of the input network as training set. One of the most important parameters of our method is α. Figure 2(a) shows the best value of α while other parameters set to default values. The effect of different size of representation vector is illustrated in figure 2(b). When Micro-f1 reaches to 128, the curve is saturated. Of course, as it stated in [22], µ and *w* can also affect on representation size.

We have shown the optimal value of µ in figure 2(c). When the number of walks for a node is increased, we are able to gather more information about that node. This would lead to more coverage of nodes' neighborhood. Although, after the number of walks reaches 40, the curve is saturated for BlogCatalog and there is no difference in Micro-f1 when we increase µ more than 40.

The Skip-gram model uses a window to extract the relationship of words that are close to each other in a document. We use the same window to relate the nodes that are located in a path. As figure 2(d) indicates, by increasing *w*, less local information about the nodes in *w* would be embedded into the representation vector. Therefore, the performance of CARE is decreased.

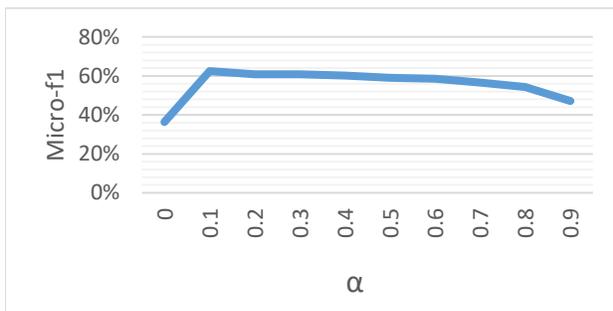
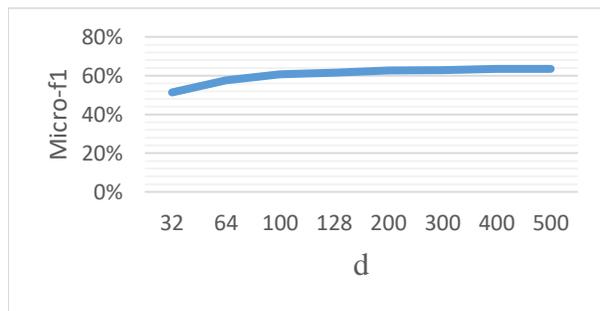

*(a) Effect of parameter α on CARE*   *(b) Effect of different dimensions on CARE*

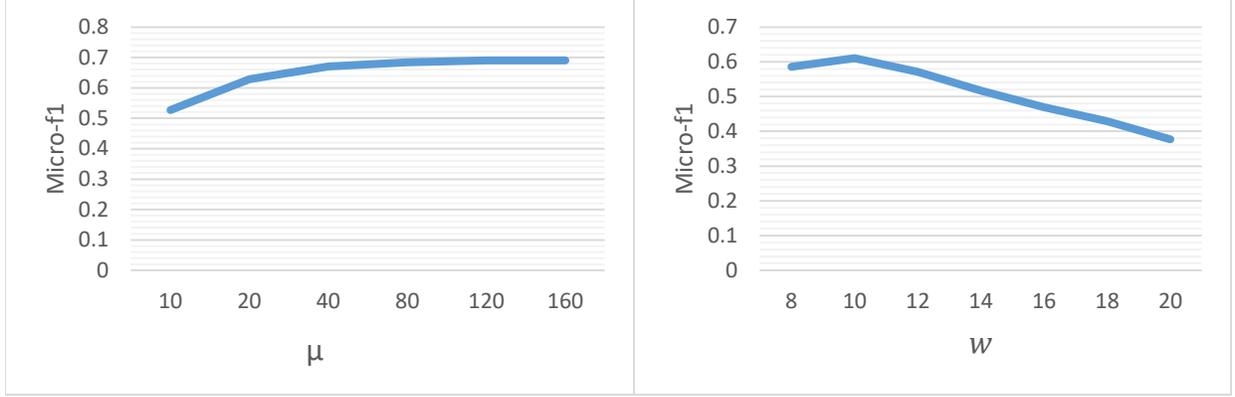

*(c) Effect of μ on CARE*  *(d) Effect of window size on CARE*

*Figure 2 effect of different parameters on CARE performance*

### 4.3. Link Prediction:

Link prediction task is a supervised learning problem that attempts to detect some future edges of the given network. We removed 50% of network edges at random to evaluate performance of CARE in link prediction. In contrast to Node2vec, we don't consider connectivity of the remained network after each edge removal. Node2vec depends on the connectivity of network. Thus, it fails to detect edges of leaf nodes in the given network. While CARE is able to detect leaf nodes' neighborhood structures using community information of them that are embedded in their representation vector though their edge was removed.

Since representation learning algorithms only generate the feature vector for each node separately, similar to [24], we also extend our algorithm by different operators to produce an edge representation for edge (u, v) such that the learned representation has the same size of the representation vectors of source and destination nodes on the edge. The operators which are considered by CARE, are defined by the following formulas [24]:

$$Hadamard: \quad [f(u) \boxdot f(v)]_i = f(u)_i * f(v)_i \qquad (9)$$

$$Average: \quad [f(u) \boxplus f(v)]_i = \frac{f(u)_i + f(v)_i}{2} \qquad (10)$$

$$Weighted - L1: \quad ||f(u) \cdot f(v)||_{1i} = |f(u)_i - f(v)_i| \qquad (11)$$

$$Weighted - L2: \quad ||f(u) \cdot f(v)||_{2i} = |f(u)_i - f(v)_i|^2 \qquad (12)$$

In the above formulas, $f(u)_i . f(v)_i$ are the i[th] features of $u$ and $v$ respectively, that are learned by the representation methods.

We confirm performance of our method in comparison to other algorithms on the datasets which their statistics are presented in Table 3.

|  | \|V\| | \|E\| |
|---|---|---|
| PPI [37] | 19706 | 390633 |
| arXiv ASTRO-PH [38] | 18772 | 198110 |

*Table 3 datasets that used for link prediction*

To learn a classifier for link prediction, we choose different training sets at random from the existing links of the network as positive edge set. We also provide negative edge set with the same size of positive edge set that contain edges that are not in the network. Then, we evaluate different algorithms with the remaining links of the network. Regression classifier is used to predict different existing and non-existing links of the network.

### 4.3.1. Experimental results:

The AUC_ROC score of our algorithm is reported in Table 4. For the link prediction task, the best value for $\alpha$ is 0.15. We have compared CARE with previous heuristic methods for the link prediction task [24]. These scores consider the number of shared immediate neighbors of nodes in different conditions as the score for each edge. Comparing the performance of CARE with heuristic methods showed about 14% improvement on arXiv dataset.

We also compare the performance of our algorithm with some representation learning algorithms, which are introduced in 4.1. Our method shows 3% gains over Node2vec algorithm on PPI dataset.

| Algorithm | arXiv | PPI |
| --- | --- | --- |
| Pref. attachments | 0.6996 | 0.6670 |
| Jaccard's Coefficient | 0.8067 | 0.7018 |
| Common neighbors | 0.8153 | 0.7142 |
| Adamic-Adar | 0.8315 | 0.7126 |
| Spectral Clustering | 0.5470 | 0.4920 |
| M-NMF | 0.9028 | 0.7318 |
| LINE | 0.8902 | 0.7249 |
| DeepWalk | 0.9340 | 0.7441 |
| Node2vec | 0.9366 | 0.7719 |
| CARE | **0.9473** | **0.7966** |

*Table 4 AUC score for different methods on link prediction (All values except for CARE come from [24] )*

Our method shows improvements on both datasets in comparison to Node2vec and DeepWalk that have the best results in representation learning algorithms. DeepWalk and Node2vec generate walks randomly and there is no information about community of nodes. Although, Node2vec attempts to consider homophily property using two controlled parameters, but these parameters unable to guarantee to preserve community information of nodes in a biased random walk. In contrast to Node2vec, CARE embeds this information into a customized random walk while we jump with probability of $\alpha$ to the nodes that are in the same community with the last node along the path.

We also investigate the effectiveness of different operators for obtaining the edges representation. The results for different operators are shown in Figure 3.

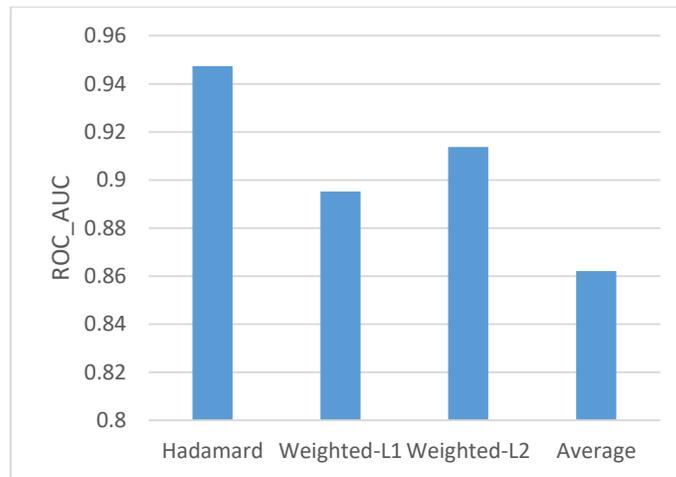

*Figure 3 – AUC_ROC Score for different operators*

As it is illustrated in Figure 3, Hadamard operator is the best choice for CARE on the link prediction task.

## 5. Conclusion:

In this paper, we have presented a novel algorithm for network embedding called CARE. To learn the representation vector of nodes, we generate some customized random walks as contextual information. In contrast to previous researches on network embedding methods, we consider both global and local neighborhood of nodes while creating paths. The Skip-gram model is used in CARE to learn the final representations of nodes. Our algorithm can embed different types of networks. The proposed method is robust to nodes addition and removal. It is scalable because it is able to generate and process customized random walks for different nodes in parallel. We have evaluated CARE on multi label classification and link prediction tasks. Experimental results on different networks show significant improvements compared to the state-of-the-art methods on network embedding.

As a part of future works, we plan to create customized random walks while we compute communities of input network to speed up CARE. We would also like to extend the proposed method to heterogeneous networks with different types of nodes and relations. In real-world networks, nodes might be in multiple communities and as another research direction; we would like to investigate the effect of various community detection algorithms such as overlapping community detection on real-world social networks using CARE. We also plan to investigate the effect of community aware random walks on Node2vec and LINE algorithms.